\documentstyle[amssymb,prl,aps,twocolumn,epsfig]{revtex}%%JN: comment out for preprint sty
\def\R  {{\rm I\kern-.15em R}}   %% set of Real numbers %%in amssymb???

\newcommand {\beq} {\begin{equation}}
\newcommand {\eeq} {\end{equation}}
\newcommand {\beqa}{\begin{eqnarray}}
\newcommand {\eeqa}{\end{eqnarray}}
\newcommand {\del} {\partial}

\newcommand {\Tr}{\mbox{Tr\,}}

\newcommand {\Pf}{\mbox{Pf}}
\newcommand {\dd}{\mbox{d}}
\newcommand {\ee}{\mbox{e}}

\begin{document}
\draft

\title{Brane world from IIB matrices}

\author{Jun Nishimura\cite{EmailJN}
        and Graziano Vernizzi\cite{EmailGV}}
\address{The Niels Bohr Institute,
Blegdamsvej 17, DK-2100 Copenhagen \O, Denmark }

\date{preprint  NBI--HE--00--31, 
hep-th/0007022; 
July, 2000
%\today %%new
     }
                                 %JN: comment next line out for preprint sty
\twocolumn[\hsize\textwidth\columnwidth\hsize\csname@twocolumnfalse\endcsname

\maketitle
\begin{abstract}
\noindent
We have recently proposed a dynamical mechanism that may realize  
a flat four-dimensional space time 
as a brane in type IIB superstring theory.
A crucial r\^{o}le is played by the phase of the chiral fermion integral
associated with the IKKT Matrix Theory, which is conjectured 
to be a nonperturbative definition of type IIB superstring theory.
We demonstrate our mechanism by studying a simplified model,
in which we find that a lower-dimensional brane indeed appears dynamically.
We also comment on some implications of our mechanism 
on model building of the brane world.
\end{abstract}
%11.25.-w Theory of fundamental strings
%11.25.Sq Nonperturbative techniques; string field theory
]   %%JN: comment out for preprint sty

\paragraph*{Introduction.---}
The idea that our four-dimensional space time is realized 
as a brane in a {\em non-compact} higher-dimensional space time
has recently attracted much attention.
Through many works during the last few years,
it is expected to provide natural resolutions to many long-standing
problems in the Standard Model.
The hierarchy problem is transmuted into a geometrical
one \cite{hierarchy}, 
and it was further argued that the exponential dependence of 
the ``warp'' factor in the extra directions
reduces the problem to a fine-tuning of order 50 \cite{RS1}.
The cosmological constant problem may also be resolved in such a 
setup \cite{Cosmolo}.
It has been argued that 
any nonzero higher-dimensional cosmological constant is absorbed
into the warp factor,
and that the four-dimensional cosmological constant
is automatically tuned to zero (or to a very small number).
A possible obstruction to the idea (as opposed to a more conventional
idea using Kaluza-Klein com\-pac\-ti\-fi\-ca\-tions) 
is that gravity may propagate
in higher dimensions and thereby contradicts 
the 4D Newton's law observed in the low energy scale.
However, the particular AdS-type background metric 
that arises naturally in such a setup
allows a normalizable zero mode of the graviton 
bound to the brane \cite{4Dgravity}.
A small correction to the 4D Newton's law due to
the continuum spectrum of massive modes is argued to be small enough to be
compatible with the experimental bound.
All these attractive features of the idea lead us to
hope that there is a natural string-theory realization of the brane
world scenario.

In Ref.~\cite{NV}, we have proposed a {\em dynamical} mechanism which
may realize a flat four-dimensional space time 
as a brane in type IIB superstring theory.
Obviously, such a mechanism should inevitably be of 
{\em nonperturbative} nature.
Indeed, our mechanism was based on the IKKT version \cite{IKKT} of 
the Matrix Theory \cite{BFSS}, namely
the IIB matrix model, which is conjectured to be a nonperturbative 
definition of type IIB superstring theory.
The model is a supersymmetric matrix model
composed by ten $N \times N$ hermitian bosonic matrices and
sixteen $N \times N$ hermitian fermionic matrices.
The space time is represented by the eigenvalues of the
bosonic matrices.
The model has manifest ten-dimensional Lorentz invariance,
where the bosonic and fermionic matrix elements transform as
a vector and a Majorana-Weyl spinor, respectively.
The integral over the fermionic matrices yields a pfaffian which is 
complex in general.
This poses a technical difficulty known as the `complex action' problem 
in studying the IIB matrix model by Monte Carlo simulation.
Monte Carlo studies incorporating only the modulus of the pfaffian
(and omitting the phase by hand) showed that the space-time becomes
isotropic in ten dimensions in the 
large-$N$ limit \cite{branched}\cite{endnote1}.
This result suggests that the phase of the pfaffian must play a crucial
r\^{o}le if a brane world naturally arises in 
the type IIB superstring theory.
The effect of the phase is
to favour configurations for which the phase becomes stationary.
Such an effect has been studied within 
a saddle-point approximation and
found to enhance lower-dimensional brane-like configurations 
considerably \cite{NV}.

In this Letter, we demonstrate our mechanism more explicitly
by studying a simplified model using Monte Carlo simulation.
In this case, we find that the dominant saddle-points
are given by configurations with only three-dimensional extent.

\paragraph*{The mechanism.---}
The IIB matrix model \cite{IKKT} is formally a
zero-volume limit of $D=10$, ${\cal N}=1$, 
pure super Yang-Mills theory.
The partition function of the IIB matrix model
(and its generalizations to $D=4$ and $D=6$) can be written as
\beq
Z_{\rm IIB} =  \int \dd A ~ \ee^{-S_{\rm b} }~Z_f [A]   \ ,
\label{original_model}
\eeq
where $S_{\rm b}= - \Tr([A_{\mu},A_{\nu}]^2)/4$,
and $\Gamma [A] = - \ln Z_f [A]$ represents the effective action induced
by integration over the fermionic matrices.
The dynamical variables $A_{\mu}$ ($\mu = 1,\cdots , D$) are 
$D$ bosonic $N \times N$ traceless hermitian matrices.
Expanding $A_{\mu}$ as $A_{\mu}=\sum_{a=1}^{N^2-1} A_{\mu}^{a} t^a$
in terms of the generators $t^a$ $(a=1, \ldots , N^2-1)$
of SU($N$), the integration measure $\dd A$ is given as
$ {\mbox{d}} A \equiv  
\prod_{\mu=1}^{D} \prod_{a=1}^{N^2-1} dA_{\mu}^{a}$.
The generators $t^a$ are normalized as $\Tr (t^a t^b) = 2 \delta _{ab}$.

The fermion integral $Z_f [A]$ is 
complex in general for $D=10$, $N\ge 4$ 
and for $D=6$, $N\ge 3$ \cite{NV}\cite{endnote0}.
We restrict ourselves to these cases in what follows.
In the $D=10$ case, the fermion integral $Z_f [A]$ is given by
the pfaffian $\Pf \, {\cal M}$, where
${\cal M}$ is a $16(N^2 -1) \times  16(N^2 -1)$ complex antisymmetric
matrix defined by
\beq
{\cal M}_{a \alpha ,  b \beta} \equiv
\Tr \left( t^a \, (\,{\cal C} \, \Gamma _\mu)_{\alpha \beta} 
[ A_\mu,t^b] \right)
\label{matrix_M} 
\eeq
regarding each of $(a\alpha)$ and $(b\beta)$ as a single index. 
Here, $\Gamma_{\mu}$ $(\mu=1, \ldots , 10)$ are
ten-dimensional Weyl-projected $16 \times 16$ gamma matrices, satisfying
${\cal C} \, \Gamma _\mu \, {\cal C}^\dag
= (\Gamma _\mu)^\top$ with ${\cal C}  = {\cal C} ^\top$ being the 
unitary charge conjugation matrix. 
Similarly in the $D=6$ case, the fermion integral $Z_f [A]$ is given by
the determinant $\det \, {\cal M}^{(6)}$, where
${\cal M}^{(6)}$ is a $4(N^2 -1) \times  4(N^2 -1)$ complex matrix 
defined by
\beq
{\cal M}^{(6)}_{a \alpha ,  b \beta} \equiv
\Tr \left( t^a \, ( \, \Gamma _\mu ^{(6)})_{\alpha \beta} 
[ A_\mu,t^b] \right)
\label{matrix_M6} 
\eeq
regarding each of $(a\alpha)$ and $(b\beta)$ as a single index. 
Here, $\Gamma_{\mu} ^{(6)}$ $(\mu=1, \ldots , 6)$ are
six-dimensional Weyl-projected $4 \times 4$ gamma matrices.

Since the fermion integral $Z_f [A]$ is a complex quantity
for the cases under consideration,
let us write it as 
$Z_f [A] = \exp (-\Gamma^{(\rm r)} - i \Gamma^{(\rm i)})$.
In Ref.~\cite{NV},
the effect of the phase $\Gamma^{(\rm i)}$
in the path integral (\ref{original_model})
has been studied using a saddle-point approximation,
whose validity has been also discussed.
The saddle-point equation for $\Gamma^{(\rm i)}$ is given by
\beq
\frac{\del \, \Gamma ^{(\rm i)}}{\del A_\mu ^a} = 0 
~~~~~
~~\forall a, \mu      \ .
\label{stationaryphase}
\eeq

It is useful to introduce the following classification 
of ``brane'' configurations
\beq
\Omega _d = \Bigl\{ \{ A_\mu \} \mid 
\exists \, n_\mu ^{(i)} (i=1,\cdots , D-d) ,
\, n_\mu ^{(i)} A_\mu = 0  \Bigr\}  \ ,
\eeq
where $n_\mu ^{(i)}$ ($i=1,\cdots , D-d$) 
are $(D-d)$ linearly independent
$D$-dimensional real vectors.
Namely, $\Omega _d$ represents a set of configurations with less than
$d$-dimensional extent.
Note that $\Omega _1 \subset \Omega _2 \subset \cdots 
\subset \Omega _{D}$, where $\Omega _{D}$ is no\-thing but the whole
configuration space of the model.
In Ref.~\cite{NV} we proved that all configurations in $\Omega _{D-2}$ 
are solutions to the saddle-point equation~(\ref{stationaryphase}).
Assuming that the configurations in $\Omega _{D-2}$ are the
dominant saddle-point configurations, we still have to
integrate over those configurations to determine 
the actual dimensionality of the space-time.
In fact, the gaussian fluctuation of the phase $\Gamma^{(\rm i)}$ 
around the saddle-points gives a huge enhancement
to the brane configurations with lower dimensionality,
and this enhancement 
cancels exactly the entropical barrier against having 
such configurations.
In the $D=10$ case,
this provides a dynamical mechanism for the possible appearance
of four-dimensional space time as a brane in ten-dimensional space time.

\paragraph*{A simplified model.---}
In order to investigate how our mechanism works,
we consider a simplified model 
de\-scri\-bing an integration over 
the saddle-point configurations.
Specifically, we consider the integral
\beq
Z = \int \dd A ~ \ee^{-\beta F[A] - \gamma f[A]} \ ,
\label{simplified_model}
\eeq
where the functions $F[A]$ and $f[A]$ are defined as
\beqa
F[A] &\equiv& \sum_{\mu=1}^{D} \sum_{a=1}^{N^2-1} 
\left(\frac{\del \, \Gamma ^{(\rm i)}}{\del A_\mu ^a} \right)^2 \ ,
\label{F_definition} \\
f[A] &\equiv& \sum_{\mu=1}^{D} \sum_{a=1}^{N^2-1} (A_{\mu}^{a})^2  \ .
\label{functionfA}
\eeqa
Since the function $F[A]$ vanishes if and only if
the configuration $\{ A_\mu \}$ satisfies the 
saddle-point equation (\ref{stationaryphase}),
the integral (\ref{simplified_model}) is dominated
by the saddle-point configurations in the large-$\beta$ limit.
The function $f[A]$ makes the integral (\ref{simplified_model}) 
convergent as long as the coefficient
$\gamma$ is fixed to be a real positive number.
In fact, the pa\-ra\-me\-ter $\gamma$ can be absorbed by an appropriate
rescaling of $A_\mu ^a$ and $\beta$.
Therefore we take $\gamma =1/2$ in what follows 
without loss of generality.
Note also that the simplified model (\ref{simplified_model})
is invariant under a SO($D$) transformation $A_\mu \mapsto \Lambda _{\mu\nu} A_\nu$,  
where $\Lambda _{\mu\nu} \in \mbox{SO}(D)$,  
and a SU($N$) transformation $A_\mu \mapsto g A_\mu g^\dagger$,
where $g \in \mbox{SU}(N)$, which are the symmetries of the original
model (\ref{original_model}).

Using the invariance of the partition function
(\ref{simplified_model}) under the change of variables
$A_{\mu} ^a \mapsto \lambda A_{\mu} ^a$,
one can obtain an exact relation
\beq
\frac{1}{2} \langle f \rangle_{\beta} 
- \beta \langle F \rangle _{\beta} =  \frac{D}{2} (N^2 - 1) \ ,
\label{exact}
\eeq
where the ensemble average
$\langle \,  \cdot \,  \rangle _{\beta}$ is defined
with the partition function (\ref{simplified_model}).
Assuming that $\langle f \rangle _{\beta}$ goes 
to a constant $c$ for $\beta \rightarrow \infty$, 
we obtain the asymptotic behavior of $\langle  F \rangle _ \beta$ 
for large $\beta$ as
\beq
\langle F \rangle_{\beta} \sim \frac{b}{\beta} \ ,
\label{Fasymptotics}
\eeq
where the coefficient $b$ is given as
$b =   c / 2  -  D (N^2 - 1)/2$.
This confirms the above claim that
the integral (\ref{simplified_model}) is dominated
by the saddle-point configurations
in the large-$\beta$ limit.

A quantity which fully characterizes the 
dimensionality $d$ of a given configuration can be 
given by the moment of inertia tensor $T$ 
defined by the $D \times D$ real symmetric matrix \cite{HNT}
\beq
T_{\mu \nu}
% = \Tr (A_{\mu} A_{\nu}) 
= \sum_{a=1}^{N^2-1} A_{\mu}^{a} A_{\nu}^{a}  \, .
\eeq
A configuration $\{ A_\mu \}$ belongs to $\Omega_d$,
if and only if the number of zero eigenvalues
of the matrix $T$ is more than or equal to $(D-d)$.
Let us denote the eigenvalues 
of the matrix $T$
as $\lambda _i$ ($i = 1, \cdots , D$),
where $\lambda_1 \ge \lambda_2 \ge \cdots \ge \lambda _{D} 
\ge 0$.
We can determine the dimensionality of the dominant
saddle-point configurations from
the ensemble average of the 
eigenvalues $\langle \lambda _ i \rangle _{\beta}$
in the $\beta \rightarrow \infty $ limit.

We address this issue by performing Monte Carlo simulation
using a Metropolis algorithm.
We create a trial configuration $\{ A_\mu ' \}$
by replacing an element $A_{\mu}^a$ of the previous configuration $\{ A_\mu \}$
with a new one generated with the
probability distribution 
$\frac{1}{\sqrt{2 \pi}}
\exp[ - \frac{1}{2}(A_\mu ^a)^2 ]$.
The trial configuration is accepted 
with the probability $\min (1, \exp (- \beta \Delta F) )$,
where $\Delta F = F[A'] - F[A]$.
This procedure is repeated for all the elements of the configuration.
The computational effort required for the above algorithm
is of order O$(N^8)$ per one sweep, which is much larger than that for 
the simulation encountered in Refs.~\cite{branched,AABHN}.
Due to this, 
results with high statistics are obtained only 
for the case of $D=6$ and $N=3$ (we have made 192,000 accepted 
updates for each $\beta \leq 384$ and 768,000 for $\beta=512$).

\begin{figure}[htbp]
  \begin{center}

 \mbox{\epsfig{file=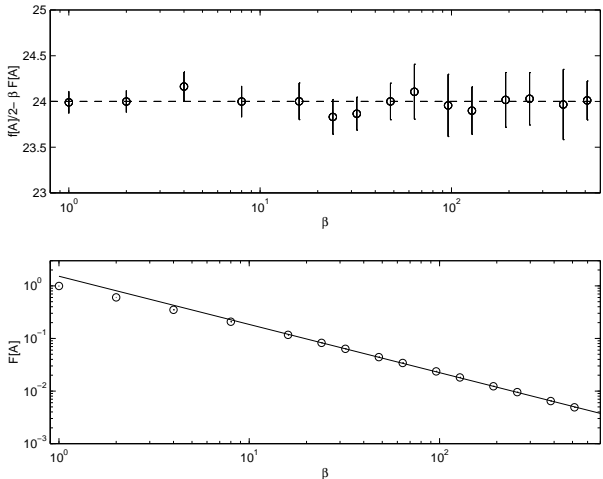,width=80mm}}
    \caption{In the upper part, we plot the left hand side of
eq.~(\ref{exact}) against $\beta$ for $D=6$ and $N=3$.
The dashed line represent the exact result $\frac{D}{2}(N^2-1)=24$.
In the lower part,
the function $\langle F[A] \rangle _\beta$ 
is plotted against $\beta$ in a log-log scale.
The straight line represents a fit to the predicted large-$\beta$
behavior $\langle F[A] \rangle _ \beta \sim b / \beta$.
}
    \label{fig:LogF}
  \end{center}
\end{figure}

\paragraph*{Results.---}
In the upper part of Fig.~\ref{fig:LogF} 
we plot the left hand side of (\ref{exact}),
which demonstrates the validity of our simulation.
In the lower part of Fig.~\ref{fig:LogF}
we plot the average $\langle F[A] \rangle _\beta$
against $\beta$ in a log-log scale.
The straight line represents the fit
of the data for $\beta \ge 16$
to the predicted large-$\beta$ behavior (\ref{Fasymptotics})
with $b=1.7(1)$.

\begin{figure}[tp]
  \begin{center}
\mbox{\epsfig{file=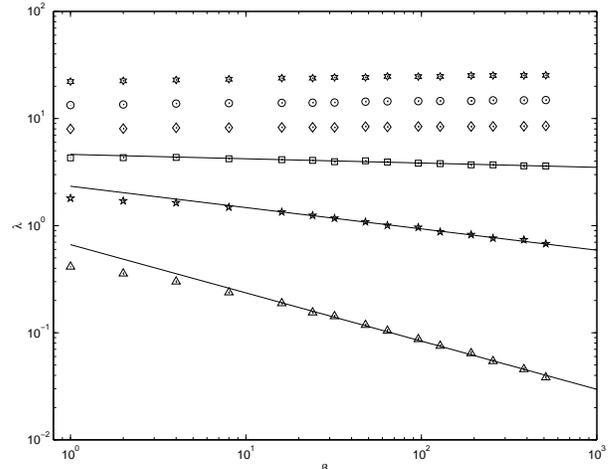,width=80mm}}
    \caption{The six eigenvalues 
$\langle \lambda_i \rangle _ \beta$ 
of the moment of inertia tensor $T$ 
are plotted against $\beta$ in a log-log scale for $D=6$ and $N=3$.
The straight lines for the smallest three eigenvalues
represent the fits to the power law behavior.}
    \label{fig:Omega6}
  \end{center}
\end{figure}

Fig.~\ref{fig:Omega6} shows
the six eigenvalues of the moment of inertia tensor $T$
as a function of $\beta$. We find that the three smallest eigenvalues 
$\langle \lambda_i\rangle_\beta$ ($i=4,5,6$)
are monotonously decreasing with a pronounced power law behavior.
Fitting the data for $\beta \ge 16$ to the power law behavior,
we extract the powers 
$-0.040(4)$,$-0.199(2)$ and $-0.450(6)$,
respectively.
Similarly, the powers for $N=4$ are extracted to be
$-0.11(1)$,$-0.26(2)$ and $-0.36(2)$.
Thus we conclude that the dominant saddle-point configurations
of the simplified model (\ref{simplified_model}) has
only three-dimensional extent.
Preliminary results for $D=10$, $N=4$ suggest that
this is the case also for $D=10$.

\paragraph*{Discussion.---}
The results presented in the previous section shows clearly that
the stationarity of the phase $\Gamma ^{(i)}$ indeed enhances 
lower-dimensional brane configurations considerably,
thus demonstrating our mechanism.
In particular, while the existence of saddle-point configurations
other than the configurations in $\Omega_{D-2}$ is not excluded,
our results suggest that such configurations, even if they exist,
can safely be neglected on statistical grounds.
Given this observation, 
the reason why we obtain 
the dimensionality `three' from the dominant saddle-point configurations
of the simplified model (\ref{simplified_model}) can be understood
analytically.
As has been done in Ref.~\cite{NV} for the IIB matrix model,
we can rewrite the $\beta \rightarrow \infty$ limit of
the simplified model (\ref{simplified_model}) as an integral
over the configurations in $\Omega_{D-2}$.
The gaussian fluctuation of the phase $\Gamma^{(\rm i)}$ 
should be taken into account by the corresponding Hesse matrix,
which is, in the present case, just the square of the one
for the IIB matrix model (or its $D=6$ version).
Due to this, the gaussian fluctuation enhances
lower-dimensional brane configurations even more strongly than
in the IIB matrix model and overwhelms the entropical barrier
against having those configurations.

The fact that the lowering of the dimensionality stops at
three instead of continuing further down can be understood as follows.
We first note that the fermion integral $Z_f [A]$ vanishes for
configurations in $\Omega_{2}$ \cite{NV}.
Therefore, the phase $\Gamma ^{(i)}$ is actually ill-defined for 
configurations in $\Omega_{2}$.
Still, we can consider configurations with $A_1$, $A_2$ being generic 
and $A_i~(i=3,\cdots ,D)$ being of order $\epsilon$.
One can easily see that the function $F[A]$ is diverging as
$\epsilon^{-2}$ for $\epsilon \rightarrow 0$.
Therefore, the $\beta$ term in (\ref{simplified_model})
suppresses configurations in $\Omega_{2}$ strongly.
(In other words, configurations in $\Omega_{2}$ are {\em not} 
saddle-point configurations, 
although $\Omega_{2}\subset \Omega_{D-2}$.)

As is clearly shown in the present work,
the enhancement occurs precisely when the space-time
becomes a {\em flat} lower-dimensional hyperplane,
which we consider as a very attractive feature of our mechanism.
Namely, our mechanism has a built-in structure in which
the brane that appears as a result of the nonperturbative string
dynamics is very likely to be flat.
According to our mechanism, scenarios with
two separated branes 
(i.e. our world and the so-called `Planck' brane as in Ref.~\cite{RS1}),
their extensions to many branes \cite{many},
and scenarios with mutually intersecting branes \cite{intersect}
seem to be unnatural.

Let us also comment on a connection of our mechanism
to the brane world scenario.
In Ref.~\cite{AIIKKT}, the IIB matrix model is expanded around
a D3-brane configuration perturbatively and 
four-dimensional noncommutative Yang-Mills theory has been 
obtained \cite{endnote4}.
The (perturbatively stable) theory,
which is obtained in this way from the IIB matrix model,
has been recently identified \cite{IIKK} with
a type IIB superstring theory in 
$\mbox{AdS}_5\times \mbox{S}^5$ with an infinite
$B$-field background.
Remarkably the metric that appears in
the corresponding supergravity solution
takes the form of Randall-Sundrum's type \cite{4Dgravity},
and thereby allows for a four-dimensional Newton's law.
We expect that brane configurations similar to 
the D3-brane configuration considered above as a {\em background}
in the IIB matrix model
should appear {\em dynamically} as a result of our mechanism.
Thus our mechanism is rather directly related to the brane world
scenario.

In the IIB matrix model, the enhancement and the entropical barrier
are exactly balanced and the actual dimensionality of the brane 
is expected to be determined as a result of large $N$ dynamics.
In this regard, we recall that
the low-energy effective theory of the
IIB matrix model has been shown to be described by a branched-polymer
like system in Ref.~\cite{AIKKT}.
There it was further argued that a typical double-tree structure
that appears in the effective theory 
might cause a collapse of the configuration to a 
lower-dimensional manifold.
Whether the actual dimensionality of the brane turns out to be
four or not can be investigated directly
by performing the integration 
over the saddle-point configurations
as formulated in Ref.~\cite{NV}.
We hope that Monte Carlo techniques 
developed in Ref.~\cite{branched} will enable us to address
such an issue in near future.

\paragraph*{Acknowledgments.---} 
We thank S.~Iso, H.~Kawai, F.R.~Klinkhamer, C.~Schmidhuber,
R.J.~Szabo and K.~Takenaga for helpful discussions.
J.N. is supported by the Japan Society for the Promotion of
Science as a Postdoctoral Fellow for Research Abroad. 
The work of G.V.~is supported by MURST (Italy) within the project 
``Fisica Teorica delle Interazioni Fondamentali''.


\begin{references}
\bibitem{endnote1} The same conclusion has been obtained
analytically for a `quenched' model, which is defined
by omitting the pfaffian completely \cite{HNT}.
\bibitem{endnote0} In the $D=4$ case, the fermion integral $Z_f [A]$
is proved to be real positive for arbitrary $N$,
which allows a standard Monte Carlo study of 
the large $N$ dynamics \cite{AABHN}.
\bibitem{endnote4} This equivalence has been understood 
at a fully nonperturbative level as a Morita equivalence of 
noncommutative gauge theories \cite{AMNS}.
\end{references}
\end{document}